\documentclass[twocolumn,preprintnumbers,amsmath,amssymb,showpacs]{revtex4}

\usepackage{graphicx}
\usepackage{dcolumn}
\usepackage{bm}
\usepackage{epsfig}

\begin{document}

\title{Relaxation Phenomena in Supercomputer Job Arrivals}

\author{Scott H. Clearwater}
\email{clearway@ix.netcom.com} \affiliation{P.O. Box 620513,
Woodside, CA 94062 USA}

\author{Stephen D. Kleban}%
\email{sdkleba@sandia.gov} \affiliation{Sandia National
Laboratories, Albuquerque, NM 87185 USA}

\date{\today}

\begin{abstract} 
We show that the distribution of supercomputer job submission
interarrival times can be understood as a relaxation process. The
process of deciding when to submit a job involves a complicated
set of interactions between the users themselves, the queuing
algorithm, the supercomputer, and a hierarchy of other decision
makers. This is analogous to the hierarchically constrained
dynamics found in glassy relaxation modelled by a stretched
exponential. Empirical supercomputer log data shows that the tails
of the distributions are well fit by a stretched exponential.
\end{abstract}


\maketitle

Today's supercomputers have thousands of processors and perform
sophisticated simulations on a wide variety of problems in
material science, structural, and thermal dynamics. Supercomputers
are an integral and enabling component in the complex system of
Big Science. Among the most powerful supercomputers are those from
the Advanced Simulation and Computing Initiative
(ASCI)\cite{ASCIref}. These machines were built for specific
purposes to primarily serve a small group of users who end up
dominating the cycles on the machine.

Supercomputers represent the largest single computing resources in
the world and they must perform over a staggering range of
conditions spanning small interactive jobs to very large jobs,
both in terms of the number of processors involved (in the
thousands) and for long time periods(on the order of a day or more
for a single run). Similar to other complex systems, the workflow
of jobs through a supercomputer system is a dynamic and
complicated cycle of phases involving submission, dispatch,
running, analysis, and resubmission. Often the ``output'' of a
phase depends critically on one or more of the other phases. For
example, the submission of a particular job at a particular time
by a particular user depends on the time the user has to spend
setting up the next run and the previous runs the user has to
analyze. These in turn depend upon when it finished running on the
machine, which depend upon when it was dispatched, which depend
upon the prioritization constraints imposed by the facility
managers via the queuing system. On top of these conditions is the
laboratory hierarchy who approve projects and above that the
governmental funding agencies and finally the elected officials
who fund the facilities.

In 1854 Kohlrausch adapted Weber's famous elasticity equation to
explain the residual charge in a Leyden jar as a function of time
and discovered the stretched exponential distribution\cite{
kohlrausch1854}, namely, that the decay time probability of the
relaxation process is given by

\begin{equation}
\label{Kohlrausch} \phi(t) = e^{(-t/\tau)^\alpha}.
\end{equation}

Since then his equation has found application in numerous
relaxation processes of complex systems in nature including
colloids, polymers, glasses, and more recently, radio emission
from galaxies, earthquakes, oilfield reserve sizes as well as
man-made phenomena such as certain market price variations and
numbers of citations\cite{ laherre98}. In this paper we will show
how job arrivals at a supercomputer can be mapped to a
hierarchical relaxation process and therefore to Kohlrausch's
result.

Heavy-tailed distributions, defined here as those that drop off
more slowly than an exponential, including the stretched
exponential and power laws, have been reported in a number of
manmade phenomena, specifically computer systems. Some examples of
heavy tail distributions in computer systems include: computer
networks both in terms of their connectivity\cite{willinger02} and
their traffic patterns\cite{willinger96bibliographical}, file
systems\cite{gribble98selfsimilarity}, video
traffic\cite{beran95}, software caches\cite{voldman83}, and the
job size distributions on a single
processor\cite{harcholbalter97}. Ultimately, these computer
systems are driven by some form of human activity interacting with
algorithms hardcoded in the hardware or programmed into the
software.

Heavy-tail distributions have important implications for both
physical and manmade systems. In particular, heavy tails indicate
a significant probability of very large events. In the case of
earthquakes it means a meaningful chance for very large and
damaging events. In the case of supercomputers it means the
possibility that the machine may become overloaded for significant
periods of time even if the average turnaround time is moderate.
Significantly, the confluence of many large jobs impinging on a
supercomputer as a consequence of heavy-tailed distributions both
in job size and interarrival time can have serious consequences on
the timeliness of the important work done at these facilities.
Thus it is important to these facilities that the implications of
these heavy-tails be characterized so that they may be taken into
account in the design of queuing algorithms and in funding
decisions for new hardware.

Despite the work in networks and single processor systems, little
is known about the scaling behavior in the largest supercomputers.
Given log data from thousands of jobs over a period of several
months we can examine these issues quantitatively.

The complicated set of conditions for determining which job gets
submitted and when is exactly the sort of conjunctive, i.e.,
multiplicative, process that is described by the stretched
exponential distribution\cite{frisch97}. More specifically, we can
think of all the hierarchy of agents interacting in getting a job
submitted to be in a discrete set of $N$ pseudospins arrayed in
different levels for each agent class. In the case of Big Science
the hierarchy is something like {\it user $\rightarrow$ project $
\rightarrow$ group $\rightarrow$ facility $\rightarrow$ laboratory
$\rightarrow$ government agency $\rightarrow$ executive and
legislative entities}.

The relaxation function, $\phi$, the probability of the system
being in a state at time $t$ is given by

\begin{equation}
\label{serialA} \phi(N, t) = 1/N\sum_{n=0}^{N}\langle
S_i(0)S_i(t)\rangle
\end{equation}

\noindent where $S_i(t)$ is the state of the $i^{th}$ pseudospin
at time $t$ and $N$ is the number of levels. In terms of an
ensemble of relaxation times we have

\begin{equation}
\label{serialB} \phi(N, t) = \sum_{n=0}^{N}w_n \exp(-t/\tau_n)
\end{equation}

\noindent where $w_n$ is the relative number of pseudospins for
level $n$. Following the arguments in \cite{ palmer84} and
\cite{klafter86}, only $\mu_n\leq N_n$ actually contribute to the
decision at the $n^{th}$-level of the hierarchy. Under this
scenario the $\mu_n$ spins in the $n^{th}$-level are free to
change only when spins in level $n-1$ have relaxed into one of
their $2^{\mu_{n-1}}$ possible states. If we ignore intralevel
correlations then

\begin{equation}
\label{relaxation} \tau_{n+1} = 2^{\mu_n}\tau_n.
\end{equation}

Defining $\tilde{\mu}_k = \mu_k\ln2$ then

\begin{equation}
\label{taus}\tau_{n+1} =
\tau_0\exp\left(\sum_{k=0}^{n}\tilde{\mu}_k\right).
\end{equation}

For a glassy relaxation, $\mu_n$ should decrease rapidly enough to
make Eq.(\ref{taus}) converge. One such condition is given by
$\mu_n = \mu_0/n$. For the job submission hierarchy we are
considering, $\mu_0$ is about $100-1000 sec$, the average time
between job submissions, the next level might be weekly meetings,
a factor of ten thousand. For the highest levels in the hierarchy,
the decisions are made on a much compressed scale relative to the
next highest level. For example, the penultimate level meets
quarterly and the highest level on a  yearly scale, a difference
of only a factor of $4$.

We also need to model the branching ratio between levels, or
``span of control'' in bureaucracy parlance. We model this as

\begin{equation}
\label{branching} w_n = w_0\lambda^{-n}.
\end{equation}

Converting the sum to an integral we have

\begin{equation}
\label{stretchedexponentialA} \phi(t) = w_0
\int_{0}^{\infty}\lambda^{-n}\exp(-tn^{-\tilde{\mu}_0}/\tau_0)dn
\end{equation}

This equation cannot be solved in closed form so by the method of
steepest descent expanding around the point $n\propto t^\alpha$ we
finally obtain the desired result, Eq.(\ref{Kohlrausch}), where
$\tau$ defines a characteristic scale to the distribution
(contrast this with a power law's scale free behavior) and finally

\begin{equation}
\label{alpha} \alpha=1/(1 + \tilde{\mu}_0)
\end{equation}

\noindent is a measure of the heaviness of the tail. The smaller
the value of $\alpha$, the heavier the tail.

Another way of looking at a relaxation process is as a random walk
in a fractal space\cite{ jund00}. When the relaxation process is
described by a stretched exponential this is seen as the signature
of a fractal morphology of the configuration space at the current
temperature of the system. In this view the complex morphology of
the job submission landscape as the set of necessary steps needed
for submission fall into place is what drives the system into its
heavy-tailed relaxation. Table \ref{tab:analogy} shows an analogy
between job events and a spin relaxation process.

\begin{table*}
\caption{\label{tab:analogy}Job event analogy to a spin relaxation
process in Nature.}
\begin{ruledtabular}
\begin{tabular}{ccccc}
Process      & Spin Glass          & Job interarrival time \\
\hline

Energy Source & Heat               & Project deliverables  \\
Energy Storage& Spins              & Pending work          \\
Threshold     & Glass transition   & Job preparation\\
              & temperature        &  \\
Energy Release& Glass transition   & Job submission        \\

\end{tabular}
\end{ruledtabular}
\end{table*}

To demonstrate that supercomputer job submissions can be
understood as a stretched exponential relaxation process, we
analyzed job logs from the ASCI supercomputers ASCI-BlueMountain
(Los Alamos National Laboratory), and ASCI-BluePacific (Lawrence
Livermore National Laboratory)\cite{ASCIref}. Each lab has devised
its own method for queuing jobs based in part on the historical
political realities at each lab\cite{clearwater02}. The important
thing to keep in mind is that the queuing algorithm through its
prioritization and ``backfilling,'' (running jobs that are not
first in the queue but can run now without slowing down the first
job in the queue) acts to alter the order that jobs were submitted
and thus when they will be dispatched, run, and finally analyzed,
all affecting the next job to be submitted and thus the
interarrival submission times.

For all the analysis shown below we tried to fit other
distributions such as the exponential, lognormal, and power law
functions, but none provided as good a fit and over such a long
range as the stretched exponential. Qualitatively, the exponential
fell off more rapidly than the data and the power law not fast
enough. The lognormal fit well for smaller values, but did poorly
at larger values, as one would expect from its functional form.
Intuitively, we might expect the stretched exponential to be
applicable and fill in this intermediate range with a moderately
heavy tail and a characteristic scale.

\begin{figure}[tbp]
\epsfig{file=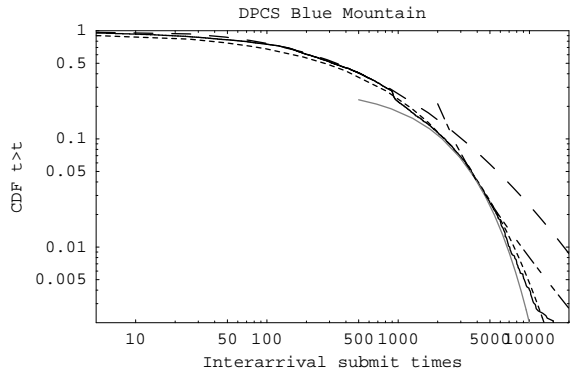,width=8cm} \caption{The
cumulative distribution function for the time between job
submissions. The average is 879s and the maximum is 161,311s. The
solid curve is the data, the dashed are from a stretched
exponential fit. The long dashed curve is a best fit to a
lognormal, the gray curve is the best fit to an exponential, and
the dot-dash is a power law over a restricted region.}
\label{fig:LSFCDFArrivalSE}
\end{figure}

Blue Mountain at Los Alamos has 5418 processors in its large
partition.There were 8171 jobs in this sample taken over a period
of 83 days. The distribution of interarrival times is shown in
Fig.~\ref{fig:LSFCDFArrivalSE}. The best fit (short dashes) to a
stretched exponential is shown with a characteristic time of
$\tau=524$s and $\alpha=.57$. Fits to lognormal and exponential
are also shown. As can clearly be seen, only the stretched
exponential is able to model the data well over its entire range.

The results from Blue Pacific at Livermore consisted of 57,430
jobs taken over a period of 63 days. Unlike Blue Mountain, Blue
Pacific did not have any partitions and used about 1000 CPUs,
although the full machine has more. The part of Blue Pacific we
used was no longer fulfilling its primary mission to the ASCI
program and is involved in more academic research.

\begin{figure}[tbp]
\epsfig{file=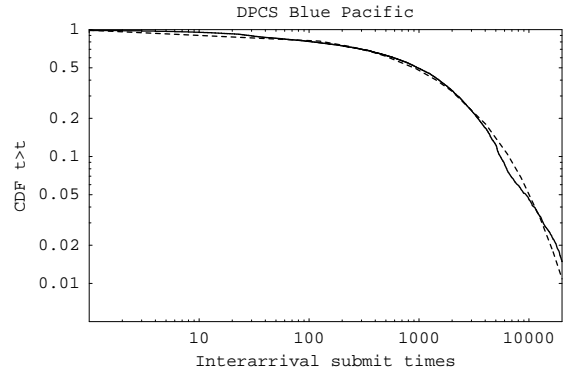,width=8cm} \caption{The
cumulative distribution function for the interarrival time between
job submissions. The solid curve is the log data, the dashed curve
is the fitted stretched exponential.} \label{fig:DPCSCDFArrivalSE}
\end{figure}

Fig.~\ref{fig:DPCSCDFArrivalSE} shows the cumulative distribution
function for interarrival times. We have truncated our fit at
10,000 seconds because beyond that time the interarrival times are
likely due to system issues and not user issues.  For example,
these events may correspond to outages in the machine or logging
errors(about 10\% of the log had bogus entries and were not used)
that could anomalously effect the very long portion of the tail.
For interarrival times up to 10,000s the parameters for the
stretched exponential are a characteristic time of $\tau=1655$s
and $\alpha=0.61$.

The parameters we found for the stretched exponential fit are
shown in Table~\ref{tab:parameters}. It is interesting to note
that for the interarrival time distribution, both LSF and DPCS
have similar exponents for large jobs, $\alpha_{LSF}=0.57$ and
$\alpha_{DPCS}=0.61$.

\begin{table}
\caption{\label{tab:parameters}Stretched exponential parameters
for LSF and DPCS data.}
\begin{ruledtabular}
\begin{tabular}{ccc}
 &\multicolumn{2}{c}{Interarrival time}\\
&$\tau$sec&$\alpha$\\ \hline

 LSF &$524$&$0.57$ \\
 DPCS&$1655$&$0.61$\\
\end{tabular}
\end{ruledtabular}
\end{table}

Our results have shown the applicability for the first time of the
stretched exponential to describing distributions from
supercomputer systems. Remarkably, the stretched exponential
provided a good fit over the entire range of values for some of
the cases we studied, spanning up to 8 orders of magnitude.

One interesting implication of the constrained hierarchical model
we are using is the relationship between levels in the hierarchy,
Eq.(\ref{relaxation}), which implies that relaxation (or response
times in our case) take much longer as one gets farther from those
doing the actual work. Interpreting the job submission process as
a relaxation phenomenon where ``barriers'' to the decision to
submit a certain sized job at a certain time must be overcome, the
exponent $\alpha$ may be understood as related to the number of
levels in a hierarchy that underlies the overall job submission
process\cite{frisch97, laherre98, palmer84}.

\begin{figure}[tbp]
\epsfig{file=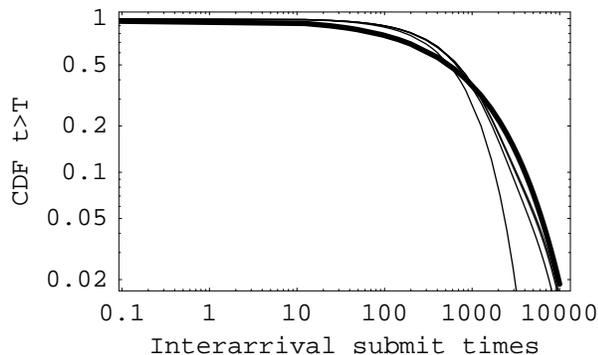,width=8cm} \caption{The cumulative
distribution function for the interarrival time between job
submissions using the fitted values(heavy curve) and the first
$\phi(N,t)$ for $N=1$ to $4$.} \label{fig:Sum}
\end{figure}

We can test the convergence of \ref{serialB} by using parameters
derived from the supercomputers. Using the empirically determined
value of $\alpha_{LSF}=0.57$, so that $\tilde{\mu}_0 = 0.75$ from
Eq.(\ref{alpha}). Values of $\mu_n < 1$ correspond to weak
constraints between levels\cite{ palmer84}, not surprising in a
scientific environment. Since we are plotting cumulative
distribution functions, the $w_n$'s themselves don't matter, but
the span of control is critical. We choose $\lambda=5$ which is a
typical span of control, $\lambda$ Eq.
(\ref{stretchedexponentialA}), in a high tech research lab. For
$\tau_0$ we use the average time between job submissions, $787$s.
We then plot $\phi(N, t)$ for $N=1$ to $4$ in Fig. \ref{fig:Sum}.
The sum converges quickly and approximates that of the exact
stretched exponential. From an organizational standpoint this
tells us that no more than 4 or so levels in the hierarchy are
having any effect on the time scale at which work gets done.

Both the supercomputers utilized in this research run under a
``Fair Share''\cite{ henry84, kay88} algorithm (user priorities
are decreased if they go over their ``share'') so it will be
interesting to see, when data becomes available, if another
queuing algorithm, such as NQS (essentially first-in first-out) at
Sandia has a similar characteristic exponent for job sizes and job
interarrival times.

The characteristic scale implied by the stretched exponential
distribution may prompt another look at some computer phenomena
previously thought to exhibit scale-free behavior. It also tells
us that the deviations from a power law are a fundamental part of
the phenomena\cite{ laherre98}. After all, as big as these
supercomputers are, they are still finite and their operators have
put in additional constraints as well to satisfy administrative
requirements, i.e., political realities. Together these
constraints act to define a characteristic size of the
distribution as well as the heaviness of the tail. For example,
the size of jobs measured in terms of number of processors and run
time was also found to be well modelled by a stretched
exponential.\cite{ clearwater02a}

In conclusion, we have shown that the interarrival time of jobs
are not exponential nor do they posses pure power-law tails, but
are somewhere in between and can be well fit by stretched
exponentials over a large and important part of their range. These
are indicative of finite scaling behaviors and have implications
for the ultimate performance of these facilities because they
relate to the frequency, and therefore the turnaround of big jobs
that are the bread and butter of the ASCI supercomputers.

This paper would not have been possible with the expertise of many
people who provided us with detailed knowledge of algorithms, data
formats, as well as providing the job log data. The authors
gratefully acknowledge the assistance of Stephany Boucher, Charles
Hales, Michael Hannah, Steve Humphreys, Moe Jette, Wilbur Johnson,
Tom Klingner, Jerry Melendez, Amy Pezzoni, Randall Rheinheimer,
Phil Salazar, Bob Wood, and Andy Yoo. Sandia is a multiprogram
laboratory operated by Sandia Corporation, a Lockheed Martin
Company, for the United States Department of Energy under Contract
DE-AC04-94AL85000.


\begin{thebibliography}{17}
\expandafter\ifx\csname natexlab\endcsname\relax\def\natexlab#1{#1}\fi
\expandafter\ifx\csname bibnamefont\endcsname\relax
  \def\bibnamefont#1{#1}\fi
\expandafter\ifx\csname bibfnamefont\endcsname\relax
  \def\bibfnamefont#1{#1}\fi
\expandafter\ifx\csname citenamefont\endcsname\relax
  \def\citenamefont#1{#1}\fi
\expandafter\ifx\csname url\endcsname\relax
  \def\url#1{\texttt{#1}}\fi
\expandafter\ifx\csname urlprefix\endcsname\relax\def\urlprefix{URL }\fi
\providecommand{\bibinfo}[2]{#2}
\providecommand{\eprint}[2][]{\url{#2}}

\bibitem[{ASC()}]{ASCIref}
\eprint{http://www.llnl.gov/asci/}.

\bibitem[{\citenamefont{Kohlrausch}(1854)}]{kohlrausch1854}
\bibinfo{author}{\bibfnamefont{R.}~\bibnamefont{Kohlrausch}},
  \bibinfo{journal}{Pogg. Ann. Phys. Chem.} \textbf{\bibinfo{volume}{91}},
  \bibinfo{pages}{179} (\bibinfo{year}{1854}).

\bibitem[{\citenamefont{Laherrere and Sornette}(1998)}]{laherre98}
\bibinfo{author}{\bibfnamefont{J.}~\bibnamefont{Laherrere}} \bibnamefont{and}
  \bibinfo{author}{\bibfnamefont{D.}~\bibnamefont{Sornette}},
  \bibinfo{journal}{European Physics Journal B} \textbf{\bibinfo{volume}{2}},
  \bibinfo{pages}{525} (\bibinfo{year}{1998}).

\bibitem[{\citenamefont{Willinger et~al.}(2002)\citenamefont{Willinger,
  Govindan, Jamin, Paxson, and Shenker}}]{willinger02}
\bibinfo{author}{\bibfnamefont{W.}~\bibnamefont{Willinger}},
  \bibinfo{author}{\bibfnamefont{R.}~\bibnamefont{Govindan}},
  \bibinfo{author}{\bibfnamefont{S.}~\bibnamefont{Jamin}},
  \bibinfo{author}{\bibfnamefont{V.}~\bibnamefont{Paxson}}, \bibnamefont{and}
  \bibinfo{author}{\bibfnamefont{S.}~\bibnamefont{Shenker}},
  \bibinfo{journal}{Proceedings of the National Academy of Sciences}
  \textbf{\bibinfo{volume}{99}}, \bibinfo{pages}{2573} (\bibinfo{year}{2002}).

\bibitem[{\citenamefont{Willinger et~al.}(1996)\citenamefont{Willinger, Taqqu,
  and Erramilli}}]{willinger96bibliographical}
\bibinfo{author}{\bibfnamefont{W.}~\bibnamefont{Willinger}},
  \bibinfo{author}{\bibfnamefont{M.}~\bibnamefont{Taqqu}}, \bibnamefont{and}
  \bibinfo{author}{\bibfnamefont{A.}~\bibnamefont{Erramilli}}
  (\bibinfo{year}{1996}).

\bibitem[{\citenamefont{Gribble et~al.}(1998)\citenamefont{Gribble, Manku,
  Roselli, Brewer, Gibson, and Miller}}]{gribble98selfsimilarity}
\bibinfo{author}{\bibfnamefont{S.~D.} \bibnamefont{Gribble}},
  \bibinfo{author}{\bibfnamefont{G.~S.} \bibnamefont{Manku}},
  \bibinfo{author}{\bibfnamefont{D.~S.} \bibnamefont{Roselli}},
  \bibinfo{author}{\bibfnamefont{E.~A.} \bibnamefont{Brewer}},
  \bibinfo{author}{\bibfnamefont{T.~J.} \bibnamefont{Gibson}},
  \bibnamefont{and} \bibinfo{author}{\bibfnamefont{E.~L.}
  \bibnamefont{Miller}}, in \emph{\bibinfo{booktitle}{Measurement and Modeling
  of Computer Systems}} (\bibinfo{year}{1998}), pp. \bibinfo{pages}{141--150},
  \urlprefix\url{citeseer.nj.nec.com/236792.html}.

\bibitem[{\citenamefont{Beran et~al.}(1995)\citenamefont{Beran, Sherman, Taquu,
  and Willinger}}]{beran95}
\bibinfo{author}{\bibfnamefont{J.}~\bibnamefont{Beran}},
  \bibinfo{author}{\bibfnamefont{R.}~\bibnamefont{Sherman}},
  \bibinfo{author}{\bibfnamefont{M.~S.} \bibnamefont{Taquu}}, \bibnamefont{and}
  \bibinfo{author}{\bibfnamefont{W.}~\bibnamefont{Willinger}},
  \bibinfo{journal}{IEEE Transactions on Communications}
  \textbf{\bibinfo{volume}{43}} (\bibinfo{year}{1995}).

\bibitem[{\citenamefont{Voldman et~al.}(1983)\citenamefont{Voldman, Mandelbrot,
  Hoevel, and J.~Knight}}]{voldman83}
\bibinfo{author}{\bibfnamefont{J.}~\bibnamefont{Voldman}},
  \bibinfo{author}{\bibfnamefont{B.}~\bibnamefont{Mandelbrot}},
  \bibinfo{author}{\bibfnamefont{L.}~\bibnamefont{Hoevel}}, \bibnamefont{and}
  \bibinfo{author}{\bibfnamefont{P.~R.} \bibnamefont{J.~Knight}},
  \bibinfo{journal}{IBM Journal of Research and Development}
  \textbf{\bibinfo{volume}{27}}, \bibinfo{pages}{164} (\bibinfo{year}{1983}).

\bibitem[{\citenamefont{Harchol-Barter and Downey}(1997)}]{harcholbalter97}
\bibinfo{author}{\bibfnamefont{M.}~\bibnamefont{Harchol-Barter}}
  \bibnamefont{and} \bibinfo{author}{\bibfnamefont{A.}~\bibnamefont{Downey}},
  \bibinfo{journal}{ACM Transactions in Computer Systems}
  \textbf{\bibinfo{volume}{15}} (\bibinfo{year}{1997}).

\bibitem[{\citenamefont{Frisch and Sornette}(1997)}]{frisch97}
\bibinfo{author}{\bibfnamefont{U.}~\bibnamefont{Frisch}} \bibnamefont{and}
  \bibinfo{author}{\bibfnamefont{D.}~\bibnamefont{Sornette}},
  \bibinfo{journal}{Phys. I France} \textbf{\bibinfo{volume}{7}}
  (\bibinfo{year}{1997}).

\bibitem[{\citenamefont{Palmer et~al.}(1984)\citenamefont{Palmer, Stein,
  Abrahams, and Anderson}}]{palmer84}
\bibinfo{author}{\bibfnamefont{R.~G.} \bibnamefont{Palmer}},
  \bibinfo{author}{\bibfnamefont{D.~L.} \bibnamefont{Stein}},
  \bibinfo{author}{\bibfnamefont{E.}~\bibnamefont{Abrahams}}, \bibnamefont{and}
  \bibinfo{author}{\bibfnamefont{P.~W.} \bibnamefont{Anderson}},
  \bibinfo{journal}{Phys. Rev. Lett.} pp. \bibinfo{pages}{958--961}
  (\bibinfo{year}{1984}).

\bibitem[{\citenamefont{Klafter and Shlesinger}(1986)}]{klafter86}
\bibinfo{author}{\bibfnamefont{J.}~\bibnamefont{Klafter}} \bibnamefont{and}
  \bibinfo{author}{\bibfnamefont{M.~F.} \bibnamefont{Shlesinger}},
  \bibinfo{journal}{Proc. Nat. Acad. Sci.} \textbf{\bibinfo{volume}{83}},
  \bibinfo{pages}{848} (\bibinfo{year}{1986}).

\bibitem[{\citenamefont{Jund et~al.}(2001)\citenamefont{Jund, Julien, and
  Campbell}}]{jund00}
\bibinfo{author}{\bibfnamefont{P.}~\bibnamefont{Jund}},
  \bibinfo{author}{\bibfnamefont{R.}~\bibnamefont{Julien}}, \bibnamefont{and}
  \bibinfo{author}{\bibfnamefont{I.}~\bibnamefont{Campbell}},
  \bibinfo{journal}{Phys. Rev. E} \textbf{\bibinfo{volume}{63}},
  \bibinfo{pages}{036131} (\bibinfo{year}{2001}).

\bibitem[{\citenamefont{Clearwater and
  Kleban}(2002{\natexlab{a}})}]{clearwater02}
\bibinfo{author}{\bibfnamefont{S.~H.} \bibnamefont{Clearwater}}
  \bibnamefont{and} \bibinfo{author}{\bibfnamefont{S.~D.}
  \bibnamefont{Kleban}}, in \emph{\bibinfo{booktitle}{16th International
  Parallel \& Distributed Processing Symposium}}
  (\bibinfo{year}{2002}{\natexlab{a}}).

\bibitem[{\citenamefont{Henry}(1984)}]{henry84}
\bibinfo{author}{\bibfnamefont{G.~J.} \bibnamefont{Henry}},
  \bibinfo{journal}{AT\&T Bell Lab. Tech. Journal}
  \textbf{\bibinfo{volume}{63}}, \bibinfo{pages}{1845} (\bibinfo{year}{1984}).

\bibitem[{\citenamefont{Kay and Lauder}(1988)}]{kay88}
\bibinfo{author}{\bibfnamefont{J.}~\bibnamefont{Kay}} \bibnamefont{and}
  \bibinfo{author}{\bibfnamefont{P.}~\bibnamefont{Lauder}},
  \bibinfo{journal}{Comm. of the ACM} \textbf{\bibinfo{volume}{31}},
  \bibinfo{pages}{44} (\bibinfo{year}{1988}).

\bibitem[{\citenamefont{Clearwater and
  Kleban}(2002{\natexlab{b}})}]{clearwater02a}
\bibinfo{author}{\bibfnamefont{S.~H.} \bibnamefont{Clearwater}}
  \bibnamefont{and} \bibinfo{author}{\bibfnamefont{S.~D.}
  \bibnamefont{Kleban}}, \bibinfo{type}{Tech. Rep.}
  \bibinfo{number}{SAND2002-2378C}, \bibinfo{institution}{Sandia National
  Laboratories} (\bibinfo{year}{2002}{\natexlab{b}}).

\end{thebibliography}

\end{document}